\documentclass[aps,twocolumn,superscriptaddress,floatfix]{revtex4}
\usepackage[dvips]{graphicx}
\usepackage{latexsym}
\usepackage{amsmath}
\usepackage{amsfonts}
\usepackage{amssymb}
\usepackage{bm}
\usepackage{color}
\usepackage{txfonts}
\begin{document}
	\newcommand{\fig}[2]{\includegraphics[width=#1]{#2}}
	\newcommand{\la}{{\langle}}
	\newcommand{\ra}{{\rangle}}
	\newcommand{\dg}{{\dagger}}
	\newcommand{\upa}{{\uparrow}}
	\newcommand{\dna}{{\downarrow}}
	\newcommand{\ab}{{\alpha\beta}}
	\newcommand{\ias}{{i\alpha\sigma}}
	\newcommand{\ibs}{{i\beta\sigma}}
	\newcommand{\hH}{\hat{H}}
	\newcommand{\hn}{\hat{n}}
	\newcommand{\hc}{{\hat{\chi}}}
	\newcommand{\hU}{{\hat{U}}}
	\newcommand{\hV}{{\hat{V}}}
	\newcommand{\br}{{\bf r}}
	\newcommand{\bk}{{{\bf k}}}
	\newcommand{\bq}{{{\bf q}}}
	\def\gsim{~\rlap{$>$}{\lower 1.0ex\hbox{$\sim$}}}
	\setlength{\unitlength}{1mm}
	\newcommand{{\vhf}}{\chi^\text{v}_f}
	\newcommand{{\vhd}}{\chi^\text{v}_d}
	\newcommand{{\vpd}}{\Delta^\text{v}_d}
	\newcommand{{\ved}}{\epsilon^\text{v}_d}
	\newcommand{{\vved}}{\varepsilon^\text{v}_d}
	\newcommand{{\tr}}{{\rm tr}}
	\newcommand{\pprl}{Phys. Rev. Lett. \ }
	\newcommand{\pprb}{Phys. Rev. {B}}

\title {Gapless vortex bound states in superconducting topological semimetals}

\author{Yi Zhang}
\thanks{These two authors contributed equally}
\affiliation{Kavli Institute of Theoretical Sciences, University of Chinese Academy of Sciences,
	Beijing, 100190, China}
\author{Shengshan Qin} 
\thanks{These two authors contributed equally}
\affiliation{Kavli Institute of Theoretical Sciences, University of Chinese Academy of Sciences,
	Beijing, 100190, China}

\author{Kun Jiang}
\email{jiangkun@iphy.ac.cn}
\affiliation{Beijing National Laboratory for Condensed Matter Physics and Institute of Physics,
	Chinese Academy of Sciences, Beijing 100190, China}
\author{Jiangping Hu}
\email{jphu@iphy.ac.cn}
\affiliation{Beijing National Laboratory for Condensed Matter Physics and Institute of Physics,
	Chinese Academy of Sciences, Beijing 100190, China}
\affiliation{Collaborative Innovation Center of Quantum Matter,
	Beijing, China}
\affiliation{Kavli Institute of Theoretical Sciences, University of Chinese Academy of Sciences,
	Beijing, 100190, China}
\date{\today}

\begin{abstract}
We find that the vortex bound states in superconducting topological  semimetals are gapless owing to topological massless excitations in their normal states.  We demonstrate this universal result in  a variety of semimetals including Dirac and Weyl semimetals, threefold degenerate spin-1 fermions,  spin-3/2 Rarita-Schwinger-Weyl fermion semimetals and other exotic fermion semimetals. The formation of these gapless bound states are closely related to their Andreev specular reflection and propagating Andreev modes in $\pi$ phase superconductor-normal metal-superconductor (SNS) junctions.  We further demonstrate that these gapless states are  topologically protected  and can be derived from a  topological pumping process. 

\end{abstract}
\maketitle

The topological semimetals (TSMs), such as Weyl and Dirac semimetals where their low energy physics  is governed by topological gapless excitations protected by topology and symmetry \cite{armitage,bradlyn}, have attracted much interest.   Recently, by studying all space-group symmetries, new additional types of TSMs  have also been  classified \cite{bradlyn,kane,weng161,weng162,soluyanov,lv17}, including spin-1 excitation\cite{manes}, spin-3/2 Rarita-Schwinger-Weyl fermion \cite{manes,liangl,ezawa,rarita} etc.  TSMs host some fantastic properties such as Fermi arcs and chiral anomaly related transports etc\cite{armitage,wangzj12,wangzj13,tang16,liu141,liu142,xusy151,wanxg,weng,lv15,xusy152,huang15, tang17,chang17,chang18,rao,sanchez,hangli}.  
 However, these exotic properties can be very different in different types of TSMs. Until now,  there is no single  universal property associated with all TSMs. 
 
 Vortex is a topological object in real space and is  a central ingredient for a type-II superconductor (SC)  under an external magnetic field \cite{gl, abrikosov}. For a conventional Fermi liquid with full gapped pairing,  the SC pairing function forms a quantum well inside a vortex core to generate a bunch of  gapped bound states, named as the Caroli–de Gennes-Matricon (CdGM) states \cite{cdm,bardeen}.  
 In this paper,  we show that the vortex bound states in superconducting  topological semimetals differ fundamentally  from conventional Fermi liquids.   There are gapless bound states in the superconducting TSM vortices, which is a universal property for TSMs beyond Dirac and Weyl semimetals\cite{qin,coleman,yan,hosur}. These states are topologically protected due to the topological gapless excitations in their normal states.  The formation of these states  is also closely related to the Andreev specular reflection and  the propagating Andreev modes in $\pi$ phase superconductor-normal metal-superconductor (SNS) junctions.

\begin{figure*}
	\begin{center}
		\fig{7.0in}{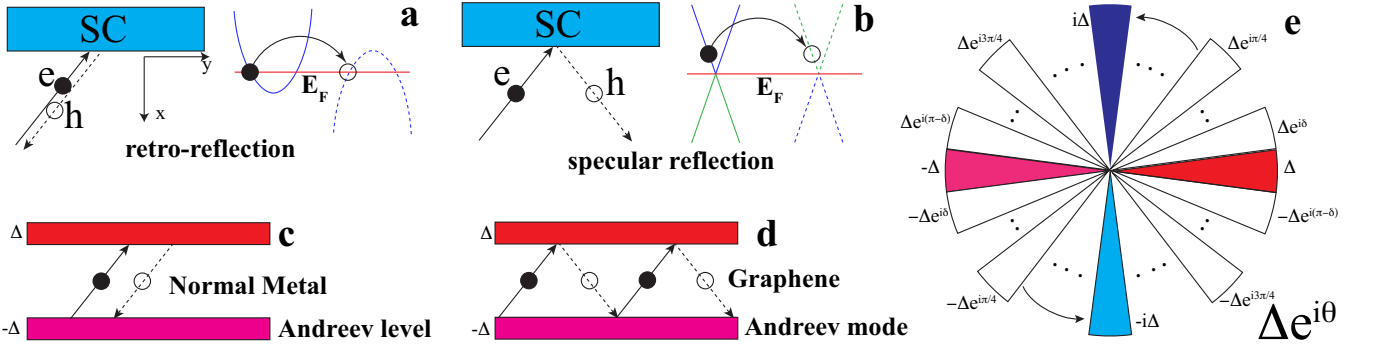}\caption{Schematic illustration of the Andreev reflection and its relation to vortex. \textbf{a}, retro-reflection in the conventional metal to SC interface, where one electron is injected with one hole with opposite velocity reflected. \textbf{b}, specular reflection in the graphene to SC interface, where the reflected hole retains its velocity along the reflection plane. \textbf{c}, Andreev level or Andreev bound state in SNS junction owing to the multiple retro-reflection at each interface. \textbf{d}, propagating Andreev mode in graphene SNS junction owing to specular reflections.
			\textbf{e}, by slicing the pairing function $\Delta e^{i\theta}$, the vortex can be mapped to infinite slices of $\pi$ junctions by taking  $\delta \rightarrow 0$. The $\pi$ junctions along the $x$ axis and $y$ axis are highlighted.
			\label{fig:fig1}}
	\end{center}
	\vskip-0.5cm
\end{figure*}

To illustrate the main idea, we consider superconducting graphene as an example\cite{graphene,geim}. In a conventional Fermi liquid, if we inject one electron into  the superconductor (SC), one hole comes out, named the Andreev retro-reflection\cite{andreev}, as shown in Fig.1a. Since the reflected hole is from the conduction hole band, its velocity is  opposite to the electron velocity from the conduction electron band. However,  in graphene, due to the topological gapless nature, the reflected hole can sit in the valence hole band when Fermi energy $E_F$ is around the Dirac point. Since the valence band has a different Fermi velocity compared with the conduction band,  the hole can retain its velocity along the reflection plane, named the specular Andreev reflection\cite{beenakker}, as shown in Fig.1b. The specular reflection leads to an unconventional behavior in the $\pi$ phase SNS junction in graphene \cite{titov}. Conventionally, the localized Andreev bound states or Andreev level can be found in the  normal metal $\pi$ junctions, as illustrated in Fig.1c \cite{sns}. On the contrary, owing to specular Andreev reflection, the graphene $\pi$ junctions show propagating Andreev modes, as shown in Fig.1d.  The physics has been generalized to the superconducting topological insulator surface states in the seminal Fu-Kane proposal \cite{fukane}, in which a tri-junction is constructed for Majorana zero mode manipulation. This tri-junction  can be viewed as a discrete analog of the superconducting topological insulator surface state vortex. 

 The above physics can be extended to a superconducting vortex. In a continuum model for graphene,  a superconducting vortex has been shown to host zero-energy bound states\cite{khaymovich,bergman}. 
 For a  vortex, the order parameter can be written as $\Delta(r)e^{i\theta}$, where $\theta$ is the phase angle centered at the vortex core. The gap value changes its sign by $180$ degree rotation. Therefore, we can slice the vortex into infinite numbers of $\pi$ junctions, as illustrated in Fig.1e. The vortex is just the superposition of all the $\pi$ junctions.
Hence, we can conjecture that if the propagating Andreev modes exist in the $\pi$ junction, there should be exotic bound states in the graphene vortex. Furthermore, it is also reasonable to argue that the specular reflection should  be a common feature for all other topological semimetals.   

\begin{figure}
	\begin{center}
		\fig{3.4in}{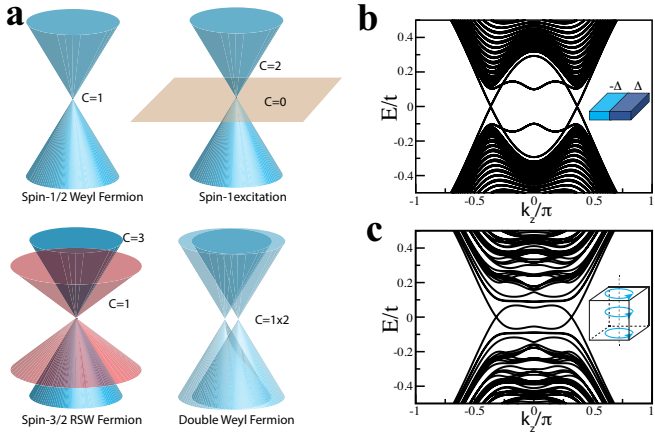}\caption{\textbf{a} Energy dispersions for multiple types of fermionic excitations including spin-1/2 Weyl fermion, spin-1 excitation, spin-3/2 RSW fermion and double  Weyl fermion. Chern number C for each band is also labeled for each fermion. \textbf{b} The $\pi$ SNS junction spectrum along the $k_z$ direction for Dirac semimetal at $k_y=0$. There are gapless dispersions around the Dirac points. \textbf{c} Vortex bound state spectrum for the Dirac semimetal. A similar gapless energy dispersion is obtained.
			\label{fig:fig2}}
	\end{center}
	\vskip-0.5cm
\end{figure}
To confirm this conjecture,  we consider various new types of fermionic semimetals  in three dimensional (3D) solids. Besides the well-established spin-1/2 Weyl fermion, spin-1, spin-3/2 massless fermionic excitations and double Weyl fermion are shown to exist owing to space group symmetry \cite{bradlyn,tang17}, as illustrated in Fig.2a. Their low-energy Hamiltonian \cite{bradlyn,tang17}, to the linear order, can be written as
\begin{eqnarray}
	H=\delta \mathbf{k} \cdot \mathbf{S}
\end{eqnarray}
where $\delta \mathbf{k}=\mathbf{k}-\mathbf{k}_0$ is the momentum deviation from the crossing point $\mathbf{k}_0$  and $\mathbf{S}$ stands for the matrices for the pseudospin operators in each spin representation.
From Fig.2a, we can see that if the $E_F$ is close to the degenerate points, the specular reflection always exists. To show this, we first take the Dirac semimetal as an example.

The lattice Hamiltonian for a Dirac semimetal can be written as
\begin{eqnarray}
	H_{D}(k)&=&(m-t\cos k_x-t\cos k_y-t_3\cos k_z)\sigma_z+t\sin k_x\sigma_x s_z  \nonumber \\
	&&+t\sin k_z(\cos k_x-\cos k_y)\sigma_x s_x-t\sin k_y \sigma_y  \nonumber\\
	&& +2t\sin k_z \sin k_x \sin k_y \sigma_x s_y \ ,
\end{eqnarray}
where the Pauli matrices $\sigma_i$ and $s_i$ with $i={x,y,z}$ act in the orbital and spin spaces, respectively  \cite{qin}. We set the hopping parameters as $\{t,t_3,m\}=\{1,0.5,2.2\}$.
$H_{D}$ hosts two Dirac points  at $(0,0,\pm k_z^c)$ with $ k_z^c=\arccos 0.4$.
With an $s$-wave superconductivity, the Hamiltonian can be written as
\begin{equation}
	H_{SC}=\left(
	\begin{array}{cccc}
		H_D(k)-\mu  & \mathbf{\hat{\Delta}(\mathbf{r})} \\
		\mathbf{\hat{\Delta}^\dagger(\mathbf{r})} & -H_D^T(-k)+\mu
	\end{array}
	\right), \label{h}
\end{equation}
in the basis $\Psi_k=\left(c_{k\uparrow},c_{k\downarrow},c_{-k\uparrow}^{\dagger},c_{-k\downarrow}^{\dagger}\right)^T$, where $\mu$ is the chemical potential and $\mathbf{\hat{\Delta}(\mathbf{r})}$ is the $s$-wave pairing function with $\hat{\mathbf{\Delta}}(\mathbf{r})=\mathbf{\Delta(\mathbf{r})}i\sigma_0s_y$.

We consider a  $\pi$ junction structure with length $L$ for the Dirac semimetal along x-direction. The pairing function is defined as
\begin{eqnarray}
	\mathbf{\Delta(\mathbf{r})}=
	\begin{cases}
		\Delta_0, &  x\leq L/2  \\\
		-\Delta_0, & x > L/2.
	\end{cases}
\end{eqnarray}
For the Dirac semimetal, there are propagating modes along the $y$ and $z$ directions. As shown in Fig.2b, a 2D gapless phase for Dirac semimetal $\pi$ junction with gapless dispersion along the $k_z$ direction is obtained numerically \cite{footnote}.

For the vortex configuration,  we consider  a flux along the (0,0,1) direction ($z$-direction). The pairing function can be written as
\begin{eqnarray}
		\mathbf{\Delta(\mathbf{r})}=\Delta(r)e^{i\theta}\ ,
\end{eqnarray}
where $r=\sqrt{x^2+y^2}$ is the distance to the vortex line and $\theta$ is the polar angle and $\Delta(r)$ takes the form
\begin{equation}
 \Delta(r)=\Delta_0\Theta(r-R)
\end{equation}
with $\Theta(r)$ the step function and R is the size of the vortex.  The qualitative results do not depend on the pairing function. If we pinch all the $\pi$ junction slices into a vortex, the vortex bound states are dispersive along $k_z$ dispersion. As plotted in Fig.2c, there is a gapless vortex bound state dispersion. This gapless feature has been obtained in Dirac semimetals\cite{qin,coleman,yan} and Weyl semimetals\cite{hosur,yan}.

There is also another large class of topological semimetals with higher order massless fermions, named the chiral crystals like RhSi, CoSi, RhGe, CoGe etc\cite{tang17,rhge,chang17,chang18,bradlyn}. For the chiral crystal, there is one long Fermi arc connecting C=2, C=-2  degeneracy points, which has been experimentally confirmed in CoSi \cite{rao,sanchez}. In order to study the physics of chiral crystals, we apply  a minimal eight-band tight-binding model, which is the prototype model for RhSi (details can be found in the supplemental materials and Ref. \cite{chang17}). The band structure for this model is plotted in Fig.3a. From the band structure, we can see
there are one Chern number C=2 spin-1 excitation (red dot) around the $\Gamma$ point and one C=-2 double Weyl Fermion around the $R$ point (green dot) in the absence of SOC, which is the general feature for the chiral crystal. To study the vortex property,  a magnetic flux along the (0,0,1) direction of the crystal is inserted. The Hamiltonian for this case can be obtained by changing the $H_D(k)$ to $H_{W}(k)$, as defined in the supplementary material \cite{sm}.

As discussed above, it is interesting to see how the vortex bound states behave around these special gapless points.  We first put the chemical potential exactly at the red spin-1 excitation point. In this case, the above condition for the specular reflection is satisfied. Hence, there should be gapless bound states inside the vortex, which is numerically calculated and plotted in Fig.3b. There are two chiral vortex bands connecting the upper and lower bounds of the SC gap around the $k_z=0$, as predicted. There are also another branch of chiral gapless vortex bound bands around the $k_z=\pi$, which is related to the R point degeneracy as discussed later.
If the chemical potential is moved to the green double Weyl points, similar gapless bound states are also obtained, as shown in Fig.3c. Interestingly, when the chemical potential sits between the spin-1 and double Weyl degeneracy point, we find the gapless vortex bound states always exist.  And if chemical potential is in the range of  $-0.75 \le \mu \le 0.59$, we can always find the vortex bound states have gapless chiral modes that disperse along the $z$-direction. Moving away from this region, a gapped vortex is obtained, similar to the normal SC cases.

We can extend  the above calculation to the spin-3/2 RSW Fermions. The spin-3/2 RSW Fermion \cite{rarita,liangl,ezawa} can be obtained by introducing SOC into the chiral crystal tight-binding model\cite{chang17}. The sixfold degeneracy at the $\Gamma$ point is split into a fourfold degenerate point that describes a spin-3/2 RSW point and a two-fold degenerate Weyl point.
Upon introducing the vortex in $z$-direction, we again obtain the gapless chiral vortex bound states dispersing along $z$-direction when the chemical potential is at the RSW degenerate point as shown in Fig.3d. Owing to SOC, the two spin degenerated vortex bound bands split into four bound bands around the $\Gamma$ point. The gapless chiral vortex bound states also exist as long as the chemical potential $\mu$ lies between $\mu_{c1}$=-0.87 and $\mu_{c2}$=0.59.

\begin{figure}
	\begin{center}
		\fig{3.4in}{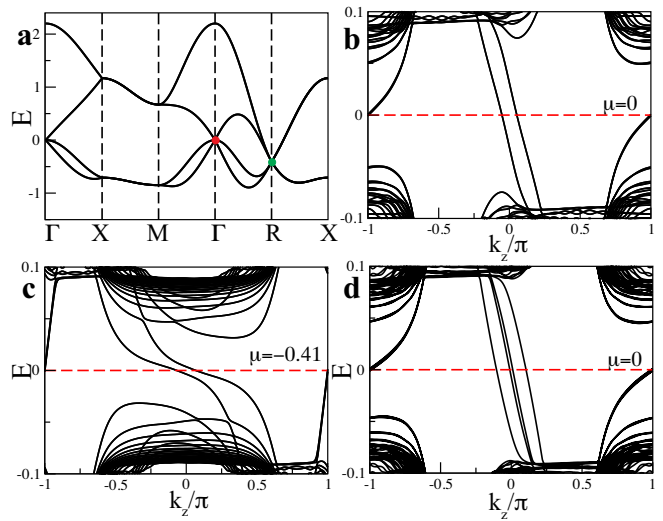}\caption{\textbf{a} Band structure for the chiral crystal eight-band model without SOC. There are one spin-1 excitation (red dot) around the $\Gamma$ point and double Weyl fermions (green dot) around the R point. \textbf{b-c} The vortex energy spectrum for \textbf{a} bands with $\mu$=0.0, -0.41, which are measured from the gapless point at $\Gamma$ point.  Two double degenerated chiral gapless vortex bound  bands locate around the $k_z=0$ while another branch of chiral bands locates around the $k_z=\pi$. \textbf{d} The vortex spectrum for the chiral crystal in the presence of SOC with $\mu$ at the RSW degenerate point. The gapless chiral bands are robust while the double degenerated chiral bands in \textbf{b} are split due to SOC.
			\label{fig:ek_soc0}}
	\end{center}
	\vskip-0.5cm
\end{figure}

\begin{figure}
	\begin{center}
		\fig{3.4in}{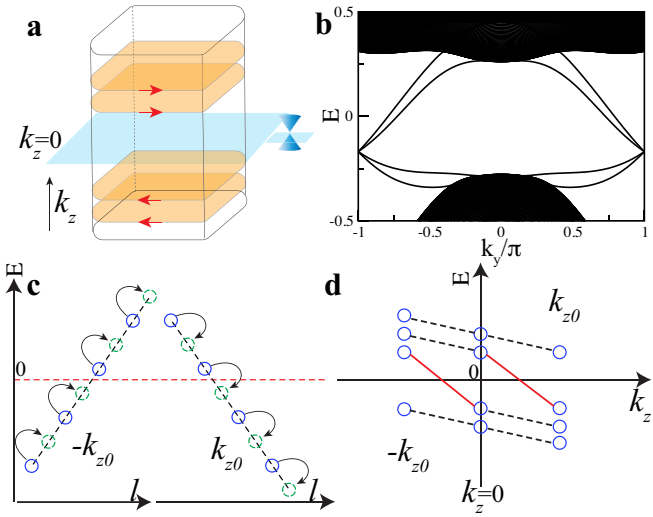}\caption{\textbf{a} Chiral edge modes above the $k_z=0$ plane and the opposite edge modes below the $k_z=0$ plane for the chiral crystal. \textbf{b} Chiral edge modes for the chiral crystal eight-band model in the presence of SOC with open boundary in the $x$-direction for a constant $k_z$ plane with $k_z=0.4\pi$, where the mode is propagating along $y$ with the chemical potential set at the 4-fold degenerate RSW point at $\Gamma$. \textbf{c} The chiral state (blue circles) energy at $\pm k_{z0}$ and their shift (green dashed circles)  under topological pumping after the flux insertion. Notice that the chiral states at $k_{z0}$ cross 0 under pumping. \textbf{d} Spectral flow for the vortex states. The spectrum for $\pm k_z$ is related to each other due to particle-hole symmetry. The bands connecting the $\pm k_{z0}$ with the $k_z=0$ plane give rise to the gapless chiral vortex bands (red lines). The connections between other states are plotted in dashed black lines.
			\label{fig:ek_soc01}}
	\end{center}
	\vskip-0.5cm
\end{figure}

All the above chiral vortex bound states are closely connected to the topological nontrivial band structures of the chiral crystal. The symmetry protected RSW fermions around the $\Gamma$ point and the Weyl fermions around $R$ point result in a long Fermi arc extending from the center to the corner in the surface Brillouin Zone. The long Fermi arc also reflects the fact that, within each $k_z$ plane ($0 < k_z < \pi$) the band structures of chiral crystal describe a 2D quantum Hall insulator with Chern number $C = 2$, and for each $-k_z$ plane the bands carry Chern number $C = -2$. The two Hamiltonians at $k_z$ and  $-k_z$ are connected by the time reversal symmetry. Accordingly, there exist two chiral edge modes on the edge of each $k_z$ plane, and the chiral edge modes carry opposite chirality for the $-k_z$ planes, as indicated in Fig.\ref{fig:ek_soc01}a and calculated in Fig.\ref{fig:ek_soc01}b.   

We can use topological flux pumping to show that there is a one-to-one correspondence between these chiral edge modes and the chiral vortex bound states in the weak pairing limit. For simplicity, we focus on one single chiral edge mode first and generalize to multiple chiral modes in the end.  We consider the vortex bound states problem in the cylinder geometry and the vortex line is parallel to the cylinder along the $z$-direction. Under this condition, in the absence of the vortex line, the chiral edge modes can be described by a series of states $H_{\text{chiral}}(k_z) = v(k_z) l_z-\mu$, with $l_z = \pm\frac{1}{2}, \pm\frac{3}{2},\ldots$ the angular momentum owing to rotation symmetry and the sign of $v(k_z)$ characterizing the chirality. Then, we can switch on a $\pi$ flux going through the vortex line. The chiral edge modes become $H_{\text{chiral}}^{\text{vortex}}(k_z) = v(k_z) l_z^\prime-\mu$, with angular momentum increasing by one-half as $l_z^\prime = l_z + \frac{1}{2}$. Namely, each chiral edge mode is shifted by half of the minimal gap and the direction of the shift is determined by the chirality. And the chiral edge modes on the edge of the $k_z$ plane and $-k_z$ plane carry opposite chirality due to time reversal symmetry. Therefore, $H_{\text{chiral}}(k_z)$ and $H_{\text{chiral}}(-k_z)$ are both shifted half of the minimal gap but in opposite directions after the $\pi$ flux is inserted.

For a chiral crystal, we can always find one insulating $k_{z0}$ plane. And each chiral edge mode in the $k_{z0}$ plane combined with its time reversal partner in the $-k_{z0}$ plane contribute to a single state which goes across the Fermi energy after the $\pi$ flux is inserted, as illustrated in Fig.4c.  This pumping process can be described by the following index $N(k_{z0}) - N(-k_{z0}) = -sgn(v(k_{z0}))$, where $N(k_z)$ is the number of states below the Fermi energy at $k_z$. For the superconducting state, an infinitesimal pairing can not change the above process, and the $\pi$ flux not only pumps an electron but also pump a hole \cite{sm}. Therefore, in the superconducting state the above index is doubled, namely $N_{\text{chiral}} = N_{\text{sc}}(k_{z0}) - N_{\text{sc}}(-k_{z0}) = -2 sgn(v(k_{z0}))$, where $N_{\text{sc}}(k_z)$ is the number of the vortex bound states with negative energy at $k_z$. Now, we can consider the spectral flow of the vortex bound states.
Owing to the particle-hole symmetry, the energy spectrum of bound states for $k_z$ is opposite to the $-k_z$. Then, the energy spectrum for $k_z=0$ is exactly particle-hole symmetric. For instance, we take  four states as one example, which leads to two positive states and two negative energy states at $k_z=0$, as shown in Fig.4d.  For $k_{z0}$, there are one positive states and three negative energy states owing to finite $N_{\text{chiral}}$. Hence, there are two chiral vortex modes connecting the $0$ to $\pm k_{z0}$ plane, as illustrated in Fig.4d. Obviously, $N_{\text{chiral}}$ is just the number of chiral vortex modes between $k_{z0}$ and $-k_{z0}$. Considering the vorticity of the vortex and the number of the chiral edge modes, the above topological index can be generalized to $N_{\text{chiral}} = -2 \eta C(k_z)$, with $\eta$ the vorticity of the vortex and $C(k_z)$ the Chern number relating to the number of chiral modes in the $k_z$ plane .

Notice that in the above analysis, we assume an insulating gap in the $k_{z0}$ plane in the normal state. For the chiral crystal, for any chemical potential $\mu$ in between $E_\Gamma$ and $E_R$, with $E_\Gamma$ and $E_R$ the energy of the degenerate point at $\Gamma$ and $R$ respectively, we can always find an insulating gap in some $k_{z0}$ plane. Therefore, for any chemical potential satisfying $E_R < \mu < E_\Gamma$, there are always four chiral vortex modes near $k_z = 0$ and four near $k_z = \pi$. 
The topological pumping process can be directly generalized to the Weyl semimetals. As illustrated in Fig.2a, the most important property for the Weyl semimetals are the Weyl points containing the spin-$1/2$ Weyl fermions with monopole charge $\text{C} = 1$. Similar to the chiral crystals, we can always find one $C\ne0$ insulating $k$ plane  with chiral edge modes, where this plane sits  between two Weyl points with opposite chirality. Based on the above analysis, the $\pi$ flux pumping gives rise to the chiral gapless vortex bound states in the superconducting Weyl semimetal \cite{hosur,yan}.

In summary, we study the gapless vortex bound states for the superconducting topological semimetals with nontrivial fermionic excitations. Owing to their fermionic excitations, the topological semimetals always host the specular Andreev reflection, which gives rise to the  propagating Andreev modes inside the TSM $\pi$ phase SNS junctions. And  the vortex can be mapped to the superposition of  $\pi$ SNS junctions. Therefore, a gapless vortex bound state can be obtained when the chemical potential sits at the degenerate points. Beyond the previously studied Dirac and Weyl semimetals gapless vortex solutions, we generalize the above discussion to the chiral crystals that hold exotic gapless fermions such as quasi spin-1 Weyl fermion and the spin-3/2 Rarita-Schwinger-Weyl fermion etc. The gapless chiral vortex bound states exist in a large region of the chemical potential. These chiral modes are topologically protected and closely related to the topologically nontrivial band structure of the chiral crystal, which can be fully understood from the topological pumping process. Our theory paves a new way to the understanding of vortex bound states in all the superconducting topological semimetals.  It is also hoped that our findings will further stimulate the investigation for the superconducting topological semimetals,  like the RhGe, which is a highly possible example for superconducting chiral crystal \cite{rhge}.

We thank Yujie Sun, Chen Fang, Li Lu for useful discussions. This work is supported by the Ministry of Science and Technology of China 973 program (Grant
No. 2017YFA0303100), Ministry of Science and Technology of China (Grant No. 2016YFA0302400), National Science Foundation of China (Grant No. NSFC-11888101, Grant
No. NSFC-11674370 and No. NSFC-11674278), and Beijing Municipal Science and Technology Commission Project
(Grant No. Z181100004218001), the Strategic Priority Research Program of Chinese Academy of Sciences (Grant
No. XDB28000000 and XDB33000000), and the Information Program of the Chinese Academy of Sciences (Grant No.
XXH13506-202). K.J. acknowledges support from the start-up grant of IOP-CAS.


\begin{thebibliography}{99}
\bibitem{footnote}
Actually there is a tiny gap at the junction spectrum  in the numerical calculation owing to the breaking of rotation symmetry.
\end{thebibliography}
\end{document}